\documentclass[%
 aip
 sd,%
 amsmath,amssymb,
 reprint,%
]{revtex4-1}

\draft
\usepackage{graphicx}
\usepackage{color}
\usepackage{dcolumn}
\usepackage{bm}

\newcommand{\rmi}{{\rm i}}

\begin{document}


\title{Resonant photonic crystals based on van der Waals heterostructures}

\author{D.R. Kazanov}
\email{kazanovdr@gmail.com}
\author{A.V. Poshakinskiy}
\author{T.V. Shubina}
\affiliation{
    Ioffe Institute, 26 Politekhnicheskaya, St Petersburg 194021, Russia
}
\date{\today}

\begin{abstract}
We propose to use 2D monolayers possessing optical gaps and high exciton oscillator strength as an element of one-dimensional resonant photonic crystals. We demonstrate that such systems are promising for the creation of effective and compact delay units. In the transition-metal-dichalcogenide-based structures where the frequencies of Bragg and exciton resonances are close, a propagating short pulse can be slowed down by few picoseconds while the pulse intensity decreases only 2\,--\,5 times. This is realized at the frequency of the ``slow'' mode situated within the stopband. The pulse retardation and attenuation can be controlled by detuning the Bragg frequency from the exciton resonance frequency.
\end{abstract}

\pacs{Valid PACS appear here}

\maketitle

Effective light pulse retardation is one of the key parts for quantum information processing \cite{MonroeC_2002}. In this particular area, it is very important to be able to control ultrafast optical signals that means working with short pulses of picosecond duration. Different Bragg structures are suitable for that purposes \cite{LiuY_2017}. A specific case of those is the resonant photonic crystals (RPCs) \cite{IvchenkoE_1994}. These periodical structures possess two types of resonances. One is the Bragg resonance that is determined by the period of the crystal. The other one is the material resonance of the optical response, e.g., exciton resonance. Overlap of these two resonances modifies optical properties of the structure and, therefore, enables control of light transmission~\cite{PrineasJ_2000}.

As a rule, one-dimensional RPCs use quantum wells (QWs) as layers with resonant optical response.
Appearance of the narrow transmission window in the stopband of such RPCs allows to delay the transmitted light pulse~\cite{YangZ_2005}. In the experiments with periodic Bragg-spaced InGaAs/GaAs QWs, the delay of light by 1.4\,ps was observed by means of cross-correlation techniques~\cite{PrineasJ_2006}. However, the damping in these structures is high ($\sim 0.99 $ of the incoming signal intensity was lost). Variation of the unit cell design, which can be either simple (single QW) or complex  (two or more QWs) \cite{IvchenkoE_2004}, allows in general to reach larger delay of the pulse with lower damping \cite{KazanovD_2017}. Importantly, it is possible to enhance the exciton resonance in RPCs composed of doubled quantum wells \cite{Bol'shakovA_2013}. This effect arises due to the formation of superradiant exciton mode that interacts with light twice as strong as the exciton in a single QW \cite{IvchenkoE_1994}.

The extreme case of a QW is a 2D semiconductor. This family includes graphene, hBN, transition-metal dichalcogenides (TMDs) monolayers and suchlike \cite{SongX_2013}. Owing to van der Waals forces one can create heterostructures from such materials \cite{GeimA_2013}. Planar photonic crystals can integrate the 2D materials to form hybrid systems of different functionality \cite{WangT_2016}. We are interested in materials with optical gap, which have exciton resonance with strong oscillator strength, like TMDs \cite{RobertC_2016a,RobertC_2016b}. These 2D semiconductors possess unique properties that differs from those in their 3D analogues \cite{MakK_2016}. In particular, the most of bulk TMDs have an indirect band gap that transforms to the direct one only when the crystal is reduced to few monolayers \cite{AroraA_2015,TongayS_2012}. The optical properties of the monolayer depend on the environment. Experiments showed that they can be markedly improved by embedding the monolayer between the layers of hBN~\cite{MancaM_2017, DufferwielS_2015}, which has indirect bandgap of $\sim 6$\,eV \cite{CassaboisG_2016}. Distinct progress in the manipulation and stacking of monolayers \cite{XiaJ_2017} stimulates the elaboration of sophisticated photonic structures based on the 2D materials.

\begin{figure}[t]
\includegraphics[scale=0.45]{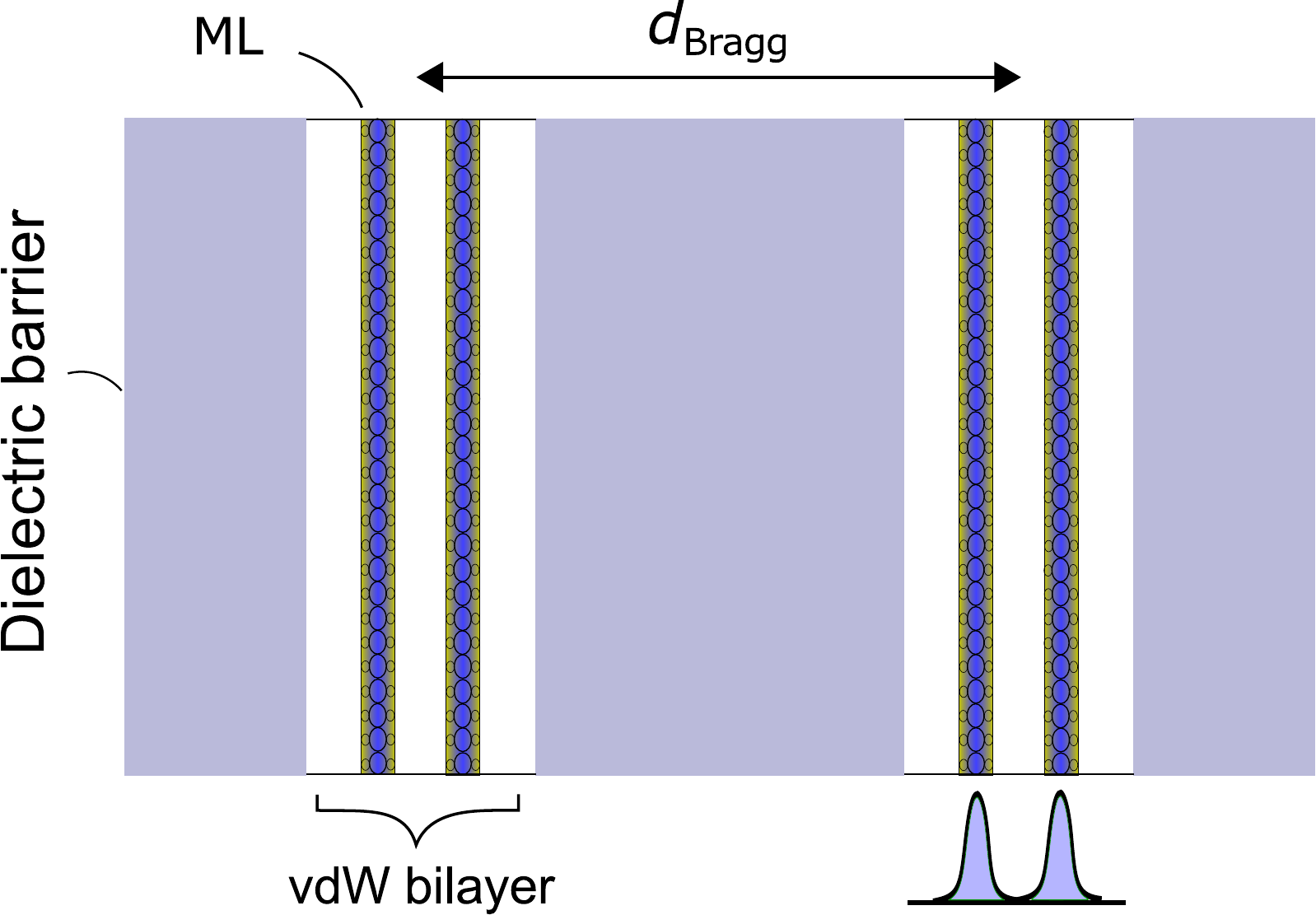}
\caption{\label{} Sketch of van der Waals Bragg structure with  period $d_{\rm{Bragg}}$. Exemplified unit cell comprises  a van der Waals bilayer sandwiched between dielectric barriers. Each 2D monolayer is surrounded by hBN shown by white color.}

\end{figure}

In this paper, we propose a novel type of resonant photonic crystals based on van der Waals monolayers. The unit cell of such RPCs comprises a TMD monolayer surrounded by dielectric barriers. The unit cell width $d_{\rm Bragg}$ determines the Bragg resonance frequency $\omega_B \approx \pi c/ (d_{\rm Bragg} \sqrt{\varepsilon})$, where $\varepsilon$ is the dielectric permittivity of the barriers.
For the sake of demonstrativeness, we assume the barriers to be SiO$_2$ or polymer with similar permittivity and choose WSe$_2$ as 2D semiconductor.
Instead of a monolayer, several layers of 2D semiconductor alternated, e.g., with hNB layers (see Fig.~1), can be used to enhance the exciton oscillator strength via superradiance effect.
We study transmission of picosecond pulses with the frequency close to the exciton resonance through these structures. Dependence of the intensity and delay of the transmitted pulse on the material parameters, detuning between Bragg and exciton resonances, and number of unit cells $N$ is analyzed, focusing on the search for the longest delay with smallest pulse attenuation.

\begin{figure}[tb]
\includegraphics[scale=0.58]{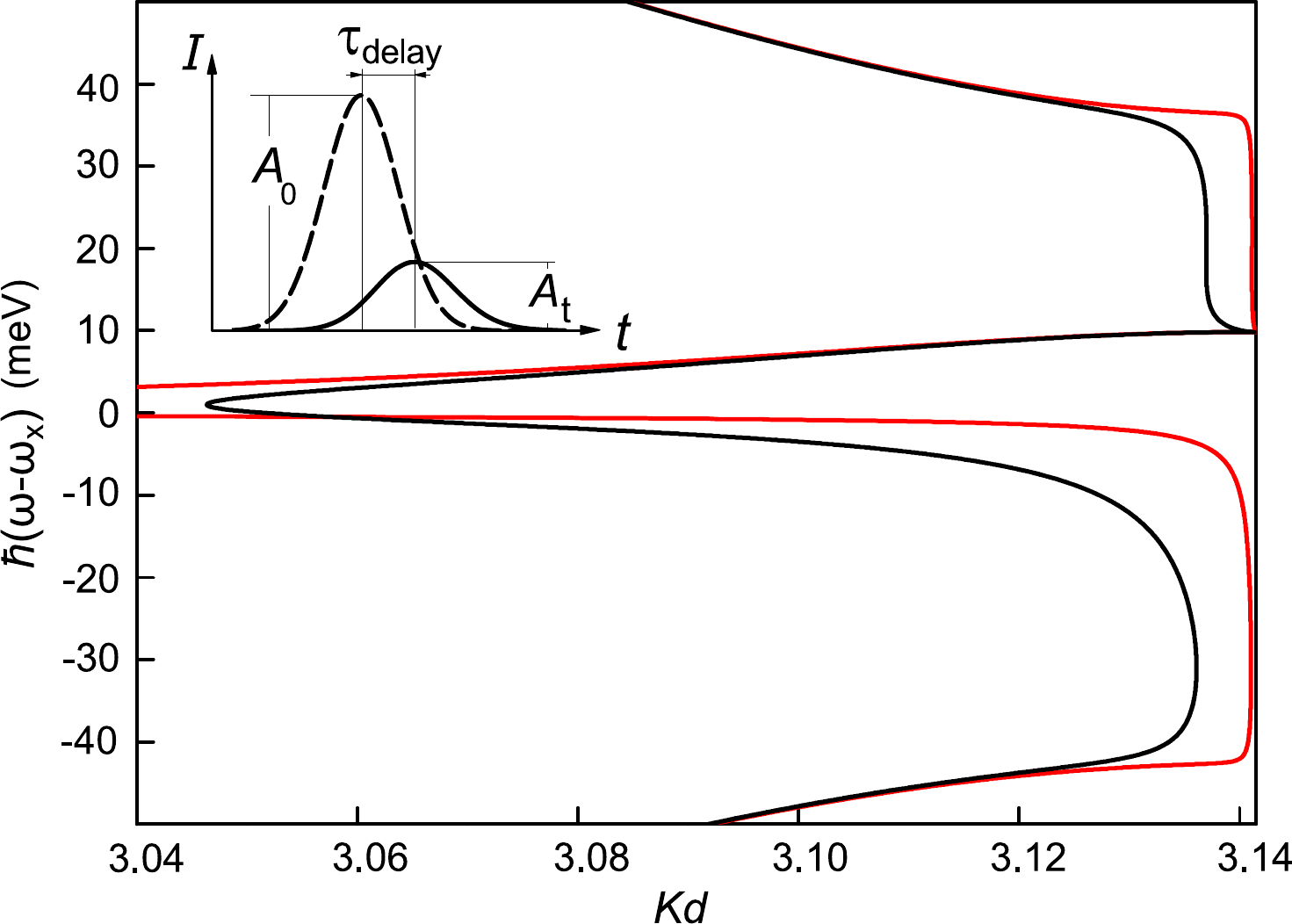}
\caption{\label{} The exciton-polariton dispersion in RPC comprizing TMD monolayers. Red and black lines correspond to non-radiative exciton width $\hbar \Gamma=0.3$ meV and $\hbar \Gamma=3$ meV, respectively; radiative width $\hbar \Gamma_0=0.6$ meV in both cases. Bragg resonance is detuned from the exciton resonance by $\hbar(\omega_B-\omega_x)=5$  meV.
The inset shows the sketch of input and transmitted pulses with respective amplitudes $A_0$ and $A_t$. The pulse delay $\tau_{\rm{delay}}$ is defined as the time difference between their maxima.
}
\end{figure}

The optical properties of a 2D monolayer in vacuum under normal incidence can be described by the transfer matrix
\begin{align}\label{eq:T2d}
\hat T_l(\omega) = \left[ \begin{array}{cc} 1+ \rmi\chi(\omega) & \rmi\chi(\omega) \\ -\rmi\chi(\omega) & 1-\rmi\chi(\omega) \end{array} \right] \,,
\end{align}
that relates the electrical field amplitudes of the right- and left-travelling waves on the right of the monolayer to those on the left of the monolayer. The form of the matrix~\eqref{eq:T2d} reflects the continuity of the electric field at the layer position. Polarizability of the 2D monolayer features the exciton resonance at the frequency $\omega_x$,
\begin{align}\label{eq:chi}
\chi(\omega) = \chi_0 - \frac{\Gamma_0}{\omega-\omega_x + \rmi\Gamma} \,.
\end{align}
Here, $\Gamma_0$ and $\Gamma$ are the radiative and non-radiative (including inhomogeneous broadening) decay rates of the exciton in the isolated monolayer. The background polarizability can be estimated as $\chi_0 \approx (\omega/c)(\varepsilon_0-1)d_0/2 \approx 0.05$, where $d_0 \approx 0.5$ nm  is the effective width of the 2D crystal and $\varepsilon_0 \approx 11.5$ is the permittivity of the bulk crystal. The 2D monolayer surrounded by material with permittivity $\varepsilon$ can be still described by the transfer matrix Eq.~\eqref{eq:T2d}, where $\chi$ should be replaced by $\bar\chi = \chi/\sqrt{\varepsilon}$. A pair of TMD mololayers separated, e.g., by hBN to avoid interlayer tunnelling has the optical response characterized by the polarizability $2\bar\chi$ and, therefore, twice higher radiative decay rate $2\Gamma_0$.

\begin{figure}[b]
\includegraphics[scale=0.7]{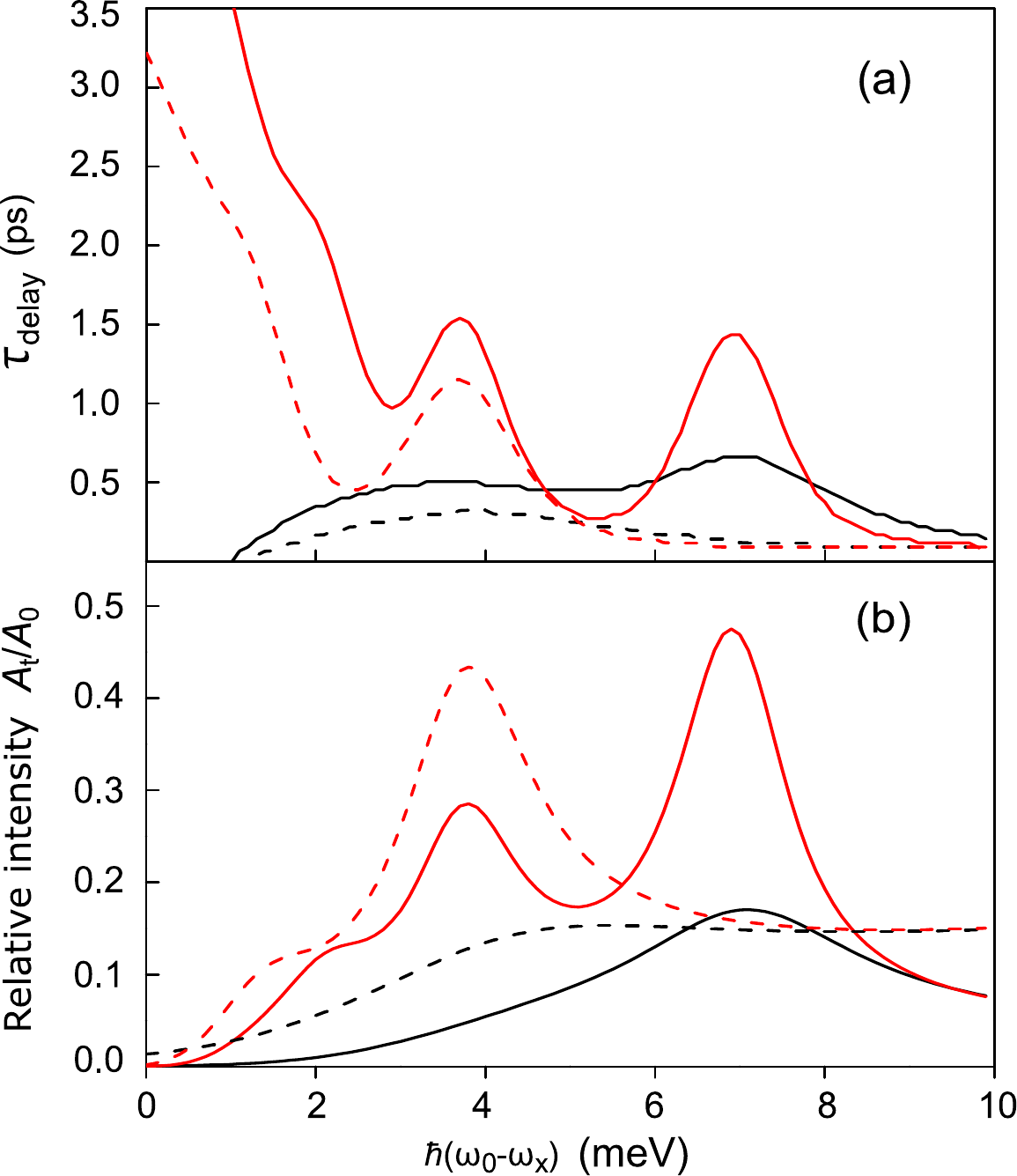}%
\caption{\label{} (a) Delay and (b) intensity of the transmitted pulse as a function of detuning of the central pulse frequency $\omega_0$  from the exciton frequency $\omega_x$. Red and black line colors correspond to $\hbar \Gamma=0.3$ meV and 3 meV, respectively; $\hbar \Gamma_0=0.6$ meV in both cases. Dashed and solid lines correspond to $N=35$ and $N=70$. }
\end{figure}

\begin{figure*}[t]
\includegraphics[scale=0.75]{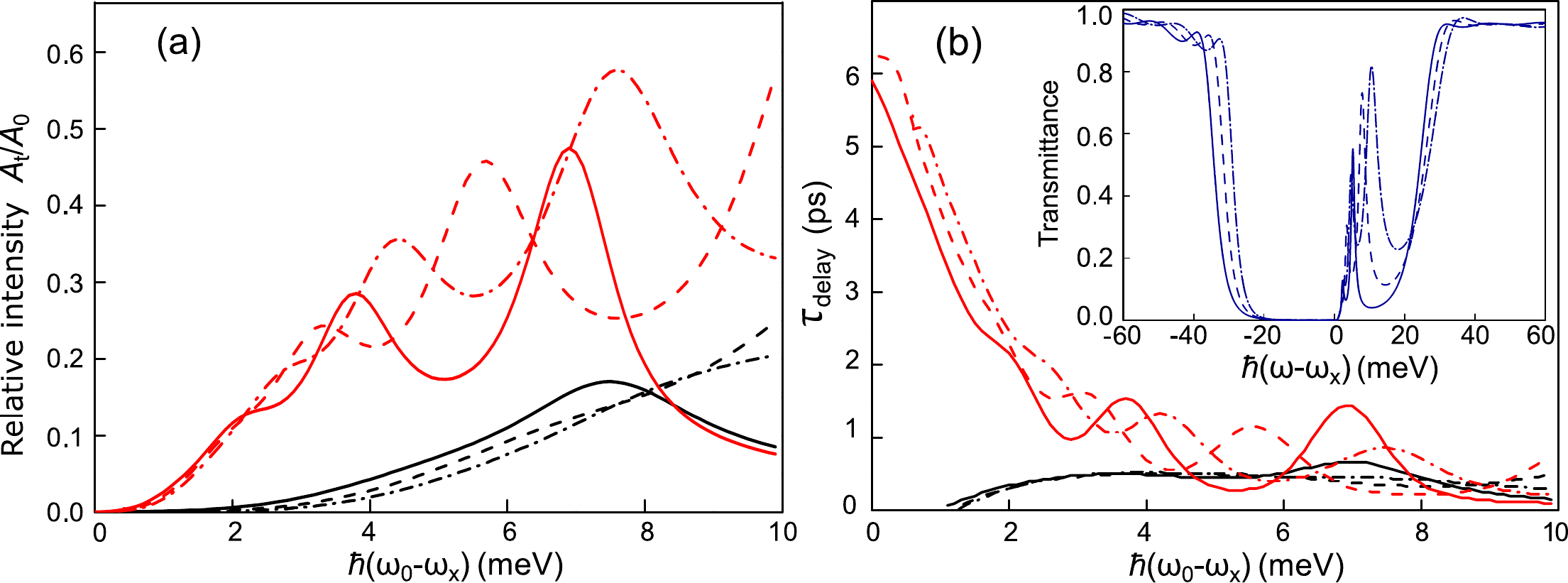}
\caption{\label{} (a) Intensity and (b) delay of the transmitted pulse as a function of central pulse frequency detuning $\omega_0-\omega_x$, calculated for different detunings between the Bragg and exciton resonances $\hbar(\omega_B-\omega_x)=$5, 10, and 15\,meV, shown by solid, dashed, and dotted lines, respectively. Red and black line colors correspond to $\hbar \Gamma=0.3$ meV and 3 meV, respectively; $\hbar \Gamma_0=0.6$ meV in both cases. The inset in panel (b) presents the transmission coefficient for the structure with the same $\omega_B-\omega_x$ detunings and $\hbar \Gamma=0.3$ meV. }
\end{figure*}

In order to estimate the realistic parameters of exciton resonance, $\omega_x$, $\Gamma_0$, and $\Gamma$, needed for the calculations, we fit the experimental reflectance spectrum measured at the low temperature in a WSe$_2$ monolayer deposited on top of a SiO$_2$ substrate, reported in \cite{WangG_2014}. Our fit gives $\hbar \omega_0=1.75$ eV, $\hbar \Gamma=3.0$ meV, $\hbar \Gamma_0=0.3$ meV. The value of $\hbar (\Gamma+\Gamma_0)$ is well consistent with the value $2.7\pm0.2$\,meV obtained by coherent spectroscopy in Ref.~\cite{MoodyG_2015}, while the value of $\hbar \Gamma_0$ is markedly less than $\hbar \Gamma_0 \sim 1.6$\,meV reported  in Refs.~\cite{MoodyG_2015, DufferwielS_2015}. In our calculations, we use  $\hbar \Gamma_0=0.6$ meV. Such value can adequately characterize both a monolayer with a strong exciton oscillator strength and a bilayer composed of a pair of layers with $\hbar \Gamma_0=0.3$ meV. The non-radiative width $\hbar \Gamma$ is chosen not only equal to 3 meV, as in the state-of-the-art structures, but also 0.3 meV, as a lower bound implied by the rapid improvement of 2D semiconductor technology.

Using transfer matrix approach we calculate the light dispersion, optical transmittance, delay and attenuation of the picosecond pulse propagating through the RPC, whose unit cells comprise the monolayers with the parameters described above.
Figure 2 shows the dispersion of exciton-polaritons in the RPC. It features a stopband caused by the Bragg resonance. The exciton resonance is located within this stopband and leads to the appearance of an additional dispersion branch which is characterized by low group velocity; therefore, termed ``slow'' mode.

The inset of Fig. 2 shows a sketch of how the input Gaussian pulse with unitary amplitude $A_0$ is modified after propagation through such a structure. The transmitted pulse is retarded by $\tau_{\rm{delay}}$ with respect to the initial pulse and its intensity is damped down to $A_t$.
Note that the group velocity determined by a derivative at the slow branch is frequency-dependent.
That means different delay values for different harmonics forming a wave packet.
However, the considered picosecond light pulse is narrower in the frequency domain ($\sim0.5$ meV) than the peculiarities of the exciton-polariton dispersion. As a result, the pulse shape remains almost unchanged after transmission through the RPC. In low-quality structures, cf. red and black lines in Fig.~2, the large non-radiative decay of the exciton broadens both the stopband and the slow branch, leading to the smaller delay.

To optimize simultaneously the delay and the intensity attenuation of the transmitted pulse we vary the detuning of the central pulse frequency $\omega_0$ from the exciton resonance with frequency $\omega_x$.
Figure 3 shows the dependencies of the delay and transmitted intensity calculated as a function of the detuning for different values of the non-radiative exciton width and number of periods. With the twice increase of the number of periods from 35 to 70, the delay pronouncedly rises; this rise is the higher the closer the pulse frequency is to the exciton resonance. However, the pulse relative intensity defined as $A_t/A_0$ drops significantly with approaching the exciton resonance. Remarkably, the variation of both the delay and the relative intensity is not monotonic but modulated by the Fabri-P\'{e}rot interference of the light trapped inside the RPC. It gives an opportunity to choose the optimal detuning $\omega_0-\omega_x$, depending on intention to obtain either larger delay or smaller attenuation, by adjustment of the central pulse frequency with an interference peak. For instance, the pulse detuned by 2 meV will provide larger delay ($\tau_{\rm{delay}}\sim2$ ps) than at 7 meV; however, at the 7-meV detuning the intensity damping is several times smaller.

We  study also how the detuning between the Bragg and exciton resonances, $\omega_B-\omega_x$, affects the pulse transmission. Figure 4 shows the intensity and delay of transmitted pulse as a function of central pulse frequency for three different values of $\omega_B-\omega_x$. With their variation from 5 meV to 15 meV, the width of the slow dispersion branch increases that leads to the shift and broadening of the interference peaks in the spectra of both the relative intensity and delay. The modification of these peaks correlates with the variation of transmission coefficient, shown in the inset of Fig.~4b. The transmittance peaks broaden as they come closer to the exciton resonance due to larger absorption.

Finally, it is worth mentioning that the manufacture of RPCs based on monolayers is, certainly, a challenge for modern nanotechnology. However, we assume that the Bragg spacing can be realized, e.g., by precise thin-film technology or by spin-coating technique developed to cover a monolayer by a polymer film \cite{LiH_2014}. Stacking of the TMD and hBN monolayers can be done by methods varying from traditional exfoliation to epitaxial growth which has rapidly developed in recent years \cite{GongY_2014, KhadkaS_2017}.

To summarize, we have studied the transmission of picosecond light pulses through the one-dimensional resonant photonic crystals of novel type that comprise atomic monolayers or bilayers as an element with resonant response. It is shown that the delay of light in the structures with the TMD monolayers can reach 2 ps with the intensity damping to 0.2\,--\,0.3 of the initial value. To optimize these parameters we propose to use the several-meV detuning between the Bragg and exciton resonances with the respective adjustment of the central pulse frequency. It is shown that the spectral modulation induced by the Fabri-P\'{e}rot interference is useful to enhance the delay in transparency windows. Thus, when designing a delay unit, the right choice of the basic parameters, such as the Bragg resonance and central pulse frequencies, is required to realize the compromise between the most effective slowdown of light and least attenuation. Advanced technology of 2D semiconductors gives certain opportunities to realize such resonant photonic structures promising for creation of compact delay cells.

\begin{acknowledgments}
This study is supported by the Government of the Russian Federation (Project \# 14.W03.31.0011).
\end{acknowledgments}

\end{document}